\documentclass[10 pt, conference]{ieeeconf}
\IEEEoverridecommandlockouts

\title{\LARGE \bf
Guaranteed $\mathcal{H}_\infty$ performance analysis and controller synthesis for interconnected linear systems from noisy input-state data
}

\author{Tom R.V. Steentjes, Mircea Lazar, Paul M.J. Van den Hof
\thanks{T.R.V. Steentjes, M. Lazar and P.M.J. Van den Hof are with the Department of Electrical Engineering, Eindhoven University of Technology, The Netherlands. E-mails: \texttt{\{t.r.v.steentjes, m.lazar, p.m.j.vandenhof\}@tue.nl}}
\thanks{This work is supported by the European Research Council (ERC), Advanced Research Grant SYSDYNET, under the European Unions Horizon 2020 research and innovation programme (grant agreement No. 694504).}
}

\usepackage{amsmath}
\usepackage{amssymb}
\usepackage{mathrsfs}
\usepackage{graphicx}
\usepackage{theorem}
\usepackage{algorithm,algorithmic}
\usepackage{xcolor}
\usepackage{cite}
\usepackage{stfloats}
\usepackage[hyphens]{url}
\usepackage{hyperref}

{\theorembodyfont{\slshape}}
{\theorembodyfont{\slshape}}
{\theorembodyfont{\slshape}\newtheorem{lemma}{Lemma}[section]}
{\theorembodyfont{\upshape}}
{\theorembodyfont{\slshape}}
{\theorembodyfont{\upshape}\newtheorem{remark}{Remark}[section]}
{\theorembodyfont{\slshape}\newtheorem{proposition}{Proposition}[section]}
{\theorembodyfont{\upshape}}
{\theorembodyfont{\upshape}}

\begin{document}
\maketitle

\begin{abstract}
The increase in available data and complexity of dynamical systems has sparked the research on data-based system performance analysis and controller design. Recent approaches can guarantee performance and robust controller synthesis based on noisy input-state data of a single dynamical system. In this paper, we extend a recent data-based approach for guaranteed performance analysis to distributed analysis of interconnected linear systems. We present a new set of sufficient LMI conditions based on noisy input-state data that guarantees $\mathcal{H}_\infty$ performance and have a structure that lends itself well to distributed controller synthesis from data. Sufficient LMI conditions based on noisy data are provided for the existence of a dynamic distributed controller that achieves $\mathcal{H}_\infty$ performance. The presented approach enables scalable analysis and control of large-scale interconnected systems from noisy input-state data sets.
\end{abstract}

\section{Introduction}
Data is becoming increasingly relevant for the analysis and control of dynamical systems. The rise in complexity of systems implies that the well-known model-based approaches can become unsuitable in applications for which the mathematical modelling is tedious. Especially for interconnected systems, such as smart grids, smart buildings or industrial processes, models are not readily available and the spatial distribution or dimensionality complicates first-principles modelling. On the other hand, data is available with increased ease~\cite{reviewarticle2017}. Data can be used either indirectly by performing system identification with model-based analysis and control, or directly via data-based system analysis and controller synthesis.

Several methods have been developed for data-based system analysis controller synthesis, we refer to~\cite{hou2013} for a survey on data-based control. Some methods rely on the reference model paradigm, such as virtual reference feedback tuning \cite{bazanella2011}, iterative feedback tuning~\cite{hjalmarsson98} and optimal controller identification~\cite{campestrini2017}. Extensions for interconnected systems to data-based distributed controller synthesis include distributed virtual reference feedback tuning in the noiseless~\cite{steentjes2020}, and noisy~\cite{steentjes2021ecc} case. The latter methods are enabled by dynamic network identification~\cite{vandenhofetal2013} and identify the distributed controller directly, instead of the target plant.

A recent trend in data-based system analysis and control originates from Willems' fundamental lemma~\cite{willems2005}. Applications include data-based predictive control~\cite{coulson2019},~\cite{cortes2020}, the data-based parameterization of stabilizing state-feedback controllers~\cite{persis2020} and robust data-based state-feedback design with noisy data~\cite{berberich2020}. The data-based verification of dissipativity properties was considered in~\cite{koch2020},~\cite{koch2020provably}, which allows to determine system measures such as the $\mathcal{H}_\infty$ norm or passivity properties from data corrupted by a noise signal satisfying quadratic bounds. A similar noise description was considered in \cite{vanwaarde2020noisy}, which extends the data-based controller design results in \cite{waarde2020tac} to the noisy case. The data-based conditions in \cite{vanwaarde2020noisy} are necessary and sufficient for stabilizing state feedback synthesis, including $\mathcal{H}_2$ or $\mathcal{H}_\infty$ performance specifications.

In this paper, we extend the framework for parameterizing an unknown system using noisy data, that was considered in the two distinct papers\cite{koch2020provably} and \cite{vanwaarde2020noisy}, to the situation of interconnected systems. The main objective of this paper is to perform \emph{scalable} analysis of an (unknown) \emph{interconnected} system and synthesis of a distributed controller from finite noisy data, with performance guarantees. The analysis is enabled by considering a dual parameterization of the set $\Sigma_{\mathcal{D}}$: the set of systems that are compatible with input-state data $\mathcal{D}$ for unmeasured noise trajectories in $\mathcal{W}$. The feature of the dual parameterization is the applicability of standard (primal) conditions for unstructured \cite{scherer2001} and structured \cite{horssen2016} robust performance analysis. For an interconnected system, we consider sets $\Sigma_{\mathcal{D}}^i$ of subsystems that are compatible with the local input-state and neighbors' state data for noise that is known to be in a set $\mathcal{W}_i$. The main contributions of this paper are as follows.

\begin{itemize}
\item[\bf{M1.}] We give a parameterization of the set $\Sigma_\mathcal{D}$ (Lemma~\ref{lem:dualparam}). This is a dual parameterization with respect to the parameterization in \cite{koch2020provably}, cf. \cite{vanwaarde2020noisy}. 
\item[\bf{M2.}] We present sufficient LMI conditions for concluding dissipativity from noisy data (Proposition~\ref{prop:disdata}), which provide a dual result of \cite[Theorem~4]{koch2020provably}.
\item[\bf{M3.}] We provide a parameterization (Lemma~\ref{lem:parami}) and dual parameterization (Lemma~\ref{lem:dualparami}) of $\Sigma_\mathcal{D}^i$ for subsystems of an interconnected system.
\item[\bf{M4.}] We present structured sufficient LMI conditions based on noisy data for concluding guaranteed $\mathcal{H}_\infty$ performance of an interconnected system (Proposition~\ref{prop:disdatai}). 
\item[\bf{M5.}] We present structured sufficient LMI conditions based on noisy data for the existence of a distributed controller that yields a guaranteed $\mathcal{H}_\infty$ performance (Proposition~\ref{prop:existci}).
\end{itemize}

The remainder of this paper is organized as follows: preliminaries are presented in Section~II. In Section~III, we provide sufficient conditions for data-based performance analysis of a system from noisy data. In Section~IV, we provide parameterizations of subsystems that are compatible with the data and structured sufficient LMIs to determine $\mathcal{H}_\infty$ performance of an interconnected system from data. The extension to sufficient data-based LMIs for the existence of a distributed controller is presented in Section~V. In Section~VI, we present two examples and conclusions are summarized in Section~VII.

\subsection*{Basic nomenclature}
The integers are denoted by $\mathbb{Z}$. Given $a\in \mathbb{Z}$, $b\in \mathbb{Z}$ such that $a<b$, we denote $\mathbb{Z}_{[a:b]}:=\left\{a,a+1,\dots,b-1,b\right\}$. Let $I_n\in\mathbb{R}^{n\times n}$, or simply $I$, denote the identity matrix and $\mathbf{1}_n\in\mathbb{R}^n$, or simply $\mathbf{1}$, denote the column vector of all ones. For a subset $A\subset\mathbb{Z}$, the vertical, respectively horizontal, stacking of matrices $X_a$, $a\in\mathcal{A}$ is denoted $\operatorname{col}_{a\in \mathcal{A}} X_a$, respectively $\operatorname{row}_{a\in \mathcal{A}} X_a$. The image and kernel of a matrix $A$ are denoted $\operatorname{im} A$ and $\operatorname{ker} A$, respectively. A matrix $A_\perp$ denotes a basis matrix of $\operatorname{ker} A$. For a real symmetric matrix $X$, $X > 0$ ($X\geq 0$) denotes that $X$ is positive (semi-) definite. The orthogonal complement of a subspace $V$ of a vector space is denoted $V^\bot$. Matrices that can be inferred from symmetry are denoted by $(\star)$.
\section{Preliminaries}
In this paper, we consider interconnected systems composed of $L$ linear time-invariant systems of the form
\begin{align}
x_i(k+1)&= A_ix_i(k)+\sum_{j\in\mathcal{N}_i}A_{ij}x_j(k)+B_iu_i(k)+w_i(k),\nonumber\\
y_i(k) &= C_ix_i(k)+D_iu_i(k)\quad \text{for } \quad i=1,\,\dots,\, L,\label{eq:i}
\end{align}
where $x_i\in\mathbb{R}^{n_i}$ denotes the state, $u_i\in\mathbb{R}^{m_i}$ the input and $w_i\in\mathbb{R}^{n_i}$ is a noise signal. The set $\mathcal{N}_i:=\{j\in\mathcal{V}\,|\, (i,j)\in\mathcal{E}\}$ denotes the neighbours of system $i$, where $\mathcal{V}$ and $\mathcal{E}\subseteq \mathcal{V}\times \mathcal{V}$ denote the set of vertices and the set of non-oriented edges defining the connected graph $\mathcal{G}=(\mathcal{V},\mathcal{E})$.

We consider the following problem set up. Let there exist a true interconnected system defined by the matrices $A_i^0$, $A_{ij}^0$ and $B_i^0$, $(i,j)\in\mathcal{E}$, generating the input-state data $\{(u_i(t),\,x_i(t)),\, t=0,\,\dots,\, N\}$ for $i\in\mathcal{V}$. This data is collected in the matrices
\begin{align*}
X_i:= [x_i(0)\ \cdots\ x_i(N)], \ U_i^-:= [u_i(0)\ \cdots\ u_i(N-1)].
\end{align*}
By defining the matrices
\begin{align*}
X_i^+ &:= [x_i(1)\ \cdots\ x_i(N)],\ X_i^-:=[x_i(0)\ \cdots\ x_i(N-1)],\\
 W_i^-&:=[w_i(0)\ \cdots\ w_i(N-1)],
\end{align*}
we obtain the data equation
\begin{align} \label{eq:dati}
X_i^+=A_i^0X_i^-+\sum_{j\in\mathcal{N}_i}A_{ij}^0X_j^-+B_i^0U_i^-+ W_i^-,
\end{align}
for each $i\in\mathcal{V}$.

Consider the stacked input, state and noise variables $u:=\operatorname{col}(u_1,\dots, u_L)$, $x:=\operatorname{col}(x_1,\dots, x_L)$ and $w:=\operatorname{col}(w_1,\dots, w_L)$. Then the interconnected system \eqref{eq:i} is compactly described by
\begin{align} \label{eq:sys}
x(k+1)&=Ax(k)+Bu(k)+w(k),\\
y(k)&=Cx(k)+Du(k),\nonumber
\end{align}
with straightforward definitions for $A$, $B$, $C$ and $D$. The corresponding data equation is
\begin{align*}
X_+ = A_0X_-+B_0U_-+W_-,
\end{align*}
with the data matrices defined for system \eqref{eq:sys} as was done for each subsystem. The transfer matrix from $u$ to $y$ of \eqref{eq:sys} is $G(q):=C(qI-A)^{-1}B+D$ and the $\mathcal{H}_\infty$ norm is denoted as $\|G\|_{\mathcal{H}_\infty}$. For $\gamma>0$, we say that the interconnected system achieves $\mathcal{H}_\infty$ performance $\gamma$ if $\|G\|_{\mathcal{H}_\infty}<\gamma$.

\section{Inferring system performance\\ from noisy data}
In this section, we consider the data-based dissipativity analysis for an unstructured system. We recall a parameterization from \cite{koch2020provably} and introduce a dual parameterization of systems that are compatible with input-state data. The dual parameterization allow us to (i) derive a dual result with respect to \cite{koch2020provably} for concluding dissipativity properties from data, and (ii) extend the data-based results to structured results for interconnected systems.

Consider the system
\begin{align} \label{eq:sys0}
x(k+1) &= A_0x(k)+B_0u(k)+w(k),\\
y(k) &= Cx(k)+Du(k)
\end{align}
with collected data 
\begin{align*}
X_+ &:= [x(1)\ \cdots\ x(N)],\ X_-:=[x(0)\ \cdots\ x(N-1)],\\
 U_-&:=[u(0)\ \cdots\ u(N-1)],
\end{align*}
and noise sequence
\begin{align*}
W_-:=[w(0)\ \cdots w(N-1)].
\end{align*}
We assume that the data $(U_-,X)$ are known, while $W_-$ is unknown, but it is an element of the set
\begin{align*}
\mathcal{W}:=\left\{W\,|\, \begin{bmatrix}
W^\top\\ I
\end{bmatrix}^\top \begin{bmatrix}
Q_w & S_w\\ S_w^\top & R_w
\end{bmatrix}\begin{bmatrix}
W^\top\\ I
\end{bmatrix}\geq 0\right\},
\end{align*}
with $Q_w<0$ to ensure that $\mathcal{W}$ is bounded. No assumptions on the statistics of $w$ are made. This noise model can represent, for example, the bound $W_-W_-^\top\leq R_w$ for $Q_w=-I$ and $S_w=0$, which has the interpretation that the energy of $w$ is bounded on the interval $[0,N-1]$, or bounds on individual components $w(k)$ \cite{vanwaarde2020noisy}. The square of sample cross-covariance bounds, as considered in \cite{hakvoort95} for parameter-bounding identification, can be captured as
\begin{align*}
\frac{1}{N} W_-R_-^\top R_- W_-^\top \leq R_w,
\end{align*}
with $Q_w=-\frac{1}{N}R_-^\top R_-$ and $S_w=0$, where $R_-$ is the data matrix of an `instrumental' signal $r$. The latter bound is particularly interesting for its general nature, although $Q_w$ is not necessarily strictly negative definite; this is a topic for future research. We assume that the data are informative enough in the sense that the matrix $\operatorname{col} (X_-, U_-)$ has full row rank.

Because the noise term is unknown, there exist multiple pairs $(A,B)$ that are compatible with the data, i.e., that satisfy
\begin{align*}
X_+=AX_-+BU_-+W_- \quad \text{ with }W_-\in\mathcal{W}.
\end{align*}
The set of all pairs of system matrices that are compatible with the data is defined as
\begin{align*}
\Sigma_{\mathcal{D}}=\{(A,B)\,|\, X_+\!=AX_-\!+BU_-\!+W \text{ for some }W\!\in\!\mathcal{W}\}.
\end{align*}
We note that the true system $(A_0,B_0)\in\Sigma_\mathcal{D}$ by construction. Furthermore, in the \emph{noiseless} case ($W_-=0$), $\Sigma_\mathcal{D}$ reduces to the singleton $\{(A_0,B_0)\}$ if $\operatorname{col} (X_-, U_-)$ has full rank \cite{waarde2020tac}.

The following result from \cite{vanwaarde2020noisy}, cf. \cite{koch2020provably}, provides a parameterization of the set $\Sigma_{\mathcal{D}}$ of all systems that are compatible with the data.
\begin{lemma}[Parameterization~$\Sigma_\mathcal{D}$]\label{lem:param}
It holds that
\begin{align*}
\Sigma_{\mathcal{D}}=\{(A,B)\,|\, \begin{bmatrix}
-A^\top\\ -B^\top \\ I
\end{bmatrix}^\top \begin{bmatrix}
\bar{Q}_{\mathcal{D}} & \bar{S}_{\mathcal{D}}\\ \bar{S}_{\mathcal{D}}^\top & \bar{R}_{\mathcal{D}}
\end{bmatrix}\begin{bmatrix}
-A^\top\\ -B^\top \\ I
\end{bmatrix}\geq 0\},
\end{align*}
with
\begin{align*}
\begin{bmatrix}
\bar{Q}_{\mathcal{D}} & \bar{S}_{\mathcal{D}}\\ \bar{S}_{\mathcal{D}}^\top & \bar{R}_{\mathcal{D}}
\end{bmatrix}=\begin{bmatrix}
X_- & 0\\ U_- & 0\\ X_+ & I
\end{bmatrix}\begin{bmatrix}
Q_w & S_w\\ S_w^\top & R_w
\end{bmatrix}\begin{bmatrix}
X_- & 0\\ U_- & 0\\ X_+ & I
\end{bmatrix}^\top.
\end{align*}
\end{lemma}
We now present a dual parameterization of $\Sigma_\mathcal{D}$.
\begin{lemma}[Dual parameterization~$\Sigma_\mathcal{D}$] \label{lem:dualparam}
Let the matrix
\begin{align*}
\begin{bmatrix}
Q_w & S_w\\ S_w^\top & R_w
\end{bmatrix}
\end{align*}
be invertible. Then it holds that
\begin{align*}
\Sigma_{\mathcal{D}}=\{(A,B)\,|\, \begin{bmatrix}
I & 0\\ 0 & I \\ A & B
\end{bmatrix}^\top \begin{bmatrix}
{Q}_{\mathcal{D}} & {S}_{\mathcal{D}}\\ {S}_{\mathcal{D}}^\top & {R}_{\mathcal{D}}
\end{bmatrix}\begin{bmatrix}
I & 0\\ 0 & I \\ A & B
\end{bmatrix}\leq 0\},
\end{align*}
where $R_\mathcal{D}>0$ with
\begin{align*}
\begin{bmatrix}
{Q}_{\mathcal{D}} & {S}_{\mathcal{D}}\\ {S}_{\mathcal{D}}^\top & {R}_{\mathcal{D}}
\end{bmatrix}:=\begin{bmatrix}
\bar{Q}_{\mathcal{D}} & \bar{S}_{\mathcal{D}}\\ \bar{S}_{\mathcal{D}}^\top & \bar{R}_{\mathcal{D}}
\end{bmatrix}^{-1}.
\end{align*}

\begin{proof}
See Appendix~\ref{app:lemproof}.
\end{proof}
\end{lemma}

Since any system that is consistent with the data is an element of $\Sigma_\mathcal{D}$, every such system admits a representation
\begin{align*}
x(k+1) = \begin{bmatrix}
A & B
\end{bmatrix}\begin{bmatrix}
x(k)\\ u(k)
\end{bmatrix},\quad \text{ with } (A,B)\in\Sigma_\mathcal{D}.
\end{align*}
As it was shown in \cite{koch2020provably}, this uncertain system admits the following LFT representation
\begin{align*}
\begin{bmatrix} x(k+1)\\ y(k)\\ p(k)\end{bmatrix}=\begin{bmatrix}
0 & 0 & I\\ C & D & 0\\ I & 0 & 0\\ 0 & I & 0
\end{bmatrix}\begin{bmatrix}
x(k)\\ u(k)\\ l(k)
\end{bmatrix},\ l(k)=\begin{bmatrix}
A & B
\end{bmatrix}p(k),
\end{align*}
with $(A,B)\in\Sigma_\mathcal{D}$.

\begin{proposition}[Dissipativity from data]\label{prop:disdata}
If there exist a $P$ and $\alpha$ such that $P>0$, $\alpha>0$ and \eqref{eq:datLMI} hold (see next page), then
\begin{align} \label{eq:LMI}
\begin{bmatrix}
I & 0\\ A & B\\ \hline 0 & I\\ C & D
\end{bmatrix}^\top\left[\begin{array}{cc|cc}
-P & 0 & 0 & 0\\ 0 & P  & 0  &0\\ \hline 0 & 0 & -Q & -S\\ 0 & 0 & -S^\top & -R
\end{array}\right]\begin{bmatrix}
I & 0\\ A & B\\ \hline 0 & I\\ C & D
\end{bmatrix}<0
\end{align}
holds for all $(A,B)\in \Sigma_\mathcal{D}$.

\begin{proof}
Let \eqref{eq:datLMI} hold and let $M=\begin{bmatrix}
A & B
\end{bmatrix}$. By Lemma~\ref{lem:dualparam}, it holds that for $\alpha>0$,
\begin{align*}
\begin{bmatrix}
I\\ M
\end{bmatrix}^\top \begin{bmatrix}
-\alpha Q_\mathcal{D} &-\alpha S_\mathcal{D}\\ -\alpha S_\mathcal{D}^\top & -\alpha R_\mathcal{D}
\end{bmatrix}\begin{bmatrix}
I\\ M
\end{bmatrix}\geq 0
\end{align*}
for all $(A,B)\in\Sigma_\mathcal{D}$. Therefore, by the full block S-procedure~\cite{scherer2001}, it follows that \eqref{eq:LMI} holds.
\end{proof}
\end{proposition}
Inequality \eqref{eq:LMI} is the well known condition for dissipativity for a quadratic supply rate matrix $\Pi=-\left[\begin{smallmatrix} Q & S\\ S^\top & R\end{smallmatrix}\right]$. A special case of the supply rate matrix is $Q = \gamma^2 I$, $ S=0$ and $R = -I$ for $\gamma>0$. For this specific case there exists a $P>0$ so that \eqref{eq:LMI} holds if and only if the channel $u \to y$ achieves $\mathcal{H}_\infty$ performance $\gamma$.

Proposition~\ref{prop:disdata} can be seen as the dual result of Theorem~4 in \cite{koch2020provably}, where a dual version of the LMI \eqref{eq:datLMI} was derived. In our procedure, we first derive a dual parameterization of $\Sigma_\mathcal{D}$, which allows the application of standard robust control tools to the LFT representation. The parameterization from Lemma~\ref{lem:param} in \cite{koch2020provably} requires the application of the dualization lemma on the data-based LMI. A feature of the dual parameterization of $\Sigma_\mathcal{D}$ in Lemma~\ref{lem:dualparam}, is that robust analysis tools for interconnected systems can be applied \emph{mutatis mutandis}, as we will show in the next section.


    \begin{figure*}[b]
    \hrulefill
\begin{align} \label{eq:datLMI}
\begin{bmatrix}
I & 0 & 0\\ 0 & I & 0\\ \hline 0 & I & 0\\ I & 0 & 0\\ 0 & 0 & I\\ \hline 0 & 0 & I\\C &0 & D
\end{bmatrix}^\top \left[\begin{array}{cc|cc|cc}
- P & 0 & 0& 0 & 0 & 0\\0 & P & 0 & 0 & 0 & 0\\ \hline
0 & 0 & -\alpha R_\mathcal{D} & -\alpha S_\mathcal{D}^\top & 0 & 0\\ 0 & 0 & -\alpha S_\mathcal{D} & -\alpha Q_\mathcal{D} & 0 & 0\\ \hline
0 & 0 & 0 & 0 & -Q & -S\\
0 & 0 & 0 & 0 & -S^\top & -R
\end{array}\right]\begin{bmatrix}
I & 0 & 0\\ 0 & I & 0\\ \hline 0 & I & 0\\ I & 0 & 0\\ 0 & 0 & I\\ \hline 0 & 0 & I\\C &0 & D
\end{bmatrix}<0
\end{align}
    \end{figure*}

\section{Interconnected system analysis}
Let us return to the interconnected system \eqref{eq:i}. We assume that for each system $i$ the data $U_i^-$, $X_i$ and $X_j$, $j\in\mathcal{N}_i$, is available, while $W_i^-$ is unknown. For each $i\in\mathcal{V}$, the noise $W_i^-$  is assumed to be an element of the set
\begin{align*}
\mathcal{W}_i=\left\{W_i\,|\, \begin{bmatrix}
W_i^\top\\ I
\end{bmatrix}^\top\begin{bmatrix}
Q_w^i & S_w^i\\ (S_w^i)^\top & R_w^i
\end{bmatrix}\begin{bmatrix}
W_i^\top\\ I
\end{bmatrix}\geq 0\right\},
\end{align*}
with $Q_i^w<0$. We assume that the data are informative enough in the sense that the matrix $\operatorname{col} (X_i^-,X_{\mathcal{N}_i}^-, U_i^-)$ has full row rank for each $i\in\mathcal{V}$.

For each subsystem, there exist multiple tuples $(A_i,A_{\mathcal{N}_i},B_i)$ that are consistent with the data, i.e., that satisfy
\begin{align} \label{eq:datai}
X_i^+=A_iX_i^-+\sum_{j\in\mathcal{N}_i}A_{ij}X_j^-+B_iU_i^-+B_i^w W_i
\end{align}
for some $W_i\in\mathcal{W}_i$. Here, we define $A_{\mathcal{N}_i}:=\operatorname{row}_{j\in\mathcal{N}_i} A_{ij}$. Hence, for each $i\in\mathcal{V}$, the set of subsystems that are consistent with the data is
\begin{align*}
\Sigma_\mathcal{D}^i :=\{(A_i,A_{\mathcal{N}_i},B_i)\,|\, \eqref{eq:datai} \text{ holds for some } W_i\in\mathcal{W}_i\}
\end{align*}
We note that under the assumption that $W_i^-\in\mathcal{W}_i$, the true system matrices are in the set $\Sigma_\mathcal{D}^i$ by construction.

\begin{lemma}[Parameterization~$\Sigma_\mathcal{D}^i$]\label{lem:parami}
It holds that
\begin{align*}
\Sigma_{\mathcal{D}}^i=\{(A_i,A_{\mathcal{N}_i},B_i)\,|\, (\star)^\top \begin{bmatrix}
\bar{Q}_{\mathcal{D}}^i & \bar{S}_{\mathcal{D}}^i\\ (\bar{S}_{\mathcal{D}}^i)^\top & \bar{R}_{\mathcal{D}}^i
\end{bmatrix}\begin{bmatrix}
-A_i^\top\\ -A_{\mathcal{N}_i}^\top\\ -B_i^\top \\ I
\end{bmatrix}\geq 0\},
\end{align*}
with
\begin{align*}
\begin{bmatrix}
\bar{Q}_{\mathcal{D}}^i & \bar{S}_{\mathcal{D}}^i\\ (\bar{S}_{\mathcal{D}}^i)^\top & \bar{R}_{\mathcal{D}}^i
\end{bmatrix}=\begin{bmatrix}
X_i^- & 0\\ X_{\mathcal{N}_i}^- & 0\\ U_i^- & 0\\ X_i^+ & I
\end{bmatrix}\begin{bmatrix}
Q_w^i & S_w^i\\ (S_w^i)^\top & R_w^i
\end{bmatrix}\begin{bmatrix}
X_i^- & 0\\ X_{\mathcal{N}_i}^- & 0\\ U_i^- & 0\\ X_i^+ & I
\end{bmatrix}^\top\!\!.
\end{align*}
\end{lemma}

\begin{lemma}[Dual parameterization~$\Sigma_\mathcal{D}^i$] \label{lem:dualparami}
Let the matrix
\begin{align*}
\begin{bmatrix}
Q_w^i & S_w^i\\ (S_w^i)^\top & R_w^i
\end{bmatrix}
\end{align*}
be invertible. It holds that $\Sigma_{\mathcal{D}}^i$ is equal to
\begin{align*}
\{(A_i,A_{\mathcal{N}_i},B_i)\,|\, (\star)^\top\! \begin{bmatrix}
{Q}_{\mathcal{D}}^i & {S}_{\mathcal{D}}^i\\ ({S}_{\mathcal{D}}^i)^\top & {R}_{\mathcal{D}}^i
\end{bmatrix}\!\begin{bmatrix}
I & 0 & 0\\ 0 & I & 0\\ 0 & 0 & I \\ A_i & A_{\mathcal{N}_i} & B_i
\end{bmatrix}\leq 0\},
\end{align*}
where $R_\mathcal{D}^i>0$ with
\begin{align*}
\begin{bmatrix}
{Q}_{\mathcal{D}}^i & {S}_{\mathcal{D}}^i\\ ({S}_{\mathcal{D}}^i)^\top & {R}_{\mathcal{D}}^i
\end{bmatrix}:=\begin{bmatrix}
\bar{Q}_{\mathcal{D}}^i & \bar{S}_{\mathcal{D}}^i\\ (\bar{S}_{\mathcal{D}}^i)^\top & \bar{R}_{\mathcal{D}}^i
\end{bmatrix}^{-1}.
\end{align*}
\end{lemma}

We note that if any interconnected system with subsystems in $\Sigma_\mathcal{D}^i$, i.e., any interconnected system that is consistent with the data, has a certain property, then also the true interconnected system has this property. To show a property for all interconnected systems that are consistent with the data, we use the following LFT representation.

Every interconnected system with subsystems in $\Sigma_{\mathcal{D}}^i$ can be described by
\begin{align*}
\begin{bmatrix}x_i(k+1)\\ y_i(k)\\ p_i(k)\end{bmatrix}=\begin{bmatrix}
0 & 0 & 0 & I\\ C_i & 0 & D_i & 0\\ \begin{bmatrix}
I\\ 0\\ 0
\end{bmatrix} & \begin{bmatrix}
0\\ I \\ 0
\end{bmatrix} & \begin{bmatrix}
0\\ 0\\ I
\end{bmatrix} & \begin{bmatrix}
0 \\ 0\\ 0
\end{bmatrix}
\end{bmatrix}\begin{bmatrix}
x_i(k)\\ \operatorname{col}_{j\in\mathcal{N}_i} x_j(k)\\ u_i(k)\\ l_i(k)
\end{bmatrix}
\end{align*}
and $l_i(k)=\begin{bmatrix}
A_i & A_{\mathcal{N}_i} & B_i
\end{bmatrix}p_i(k)$, with $(A_i,A_{\mathcal{N}_i},B_i)\in\Sigma_\mathcal{D}^i$, $i\in\mathcal{V}$.

This LFT representation for each subsystem allows us to apply robust analysis results for interconnected systems, to conclude $\mathcal{H}_\infty$ performance for all interconnected systems that are compatible with the data. Consider the matrices $Z_i$ defined in Appendix~\ref{app:scales}.
\begin{proposition}[Performance from structured data] \label{prop:disdatai}
Let $Q_\mathcal{D}^i<0$ and $\gamma>0$. If there exist $P_i$, $Z_i$ and $\alpha_i$ so that $P_i>0$, $\alpha_i>0$ and \eqref{eq:datLMIint} (see next page) holds for all $i\in\mathcal{V}$, then all interconnected systems with subsystems $(A_i, A_{\mathcal{N}_i}, B_i)\in\Sigma_\mathcal{D}^i$, $i\in\mathcal{V}$, achieve $\mathcal{H}_\infty$ performance $\gamma$.
\end{proposition}
The proof follows by a similar argument as in Proposition~\ref{prop:disdata} and the application of \cite[Theorem~1]{horssen2016} to the LFT representation.
    \begin{figure*}[b]
    \hrulefill
\begin{align} \label{eq:datLMIint}
(\star)^\top \left[\begin{array}{cc|cc|cc|cc}
-P_i & 0 & 0 & 0 & 0 & 0 & 0 & 0\\
0 & P_i & 0 & 0& 0 & 0 & 0 & 0\\ \hline
0 & 0 & Z_i^{11}& Z_i^{12} & 0 & 0 & 0 & 0\\
0 & 0 & (Z_i^{12})^\top & Z_i^{22} & 0& 0 & 0 & 0\\ \hline
0 & 0 & 0 & 0 & -\alpha_i R_\mathcal{D}^i & -\alpha_i(S_\mathcal{D}^i)^\top & 0 & 0\\
0 & 0 & 0 & 0 & -\alpha_i S_\mathcal{D}^i & -\alpha_i Q_\mathcal{D}^i & 0 & 0\\ \hline
0 & 0 & 0 & 0  & 0 & 0 &  -\gamma^2 I & 0\\
0 & 0 & 0 & 0  & 0 & 0 & 0 & I
\end{array}\right] \begin{bmatrix}
I & 0 & 0 & 0\\
0 & 0 & I & 0\\ \hline
\mathbf{1}\otimes I & 0 & 0 & 0\\ 0 & I & 0 & 0\\ \hline
0 & 0 & I & 0\\ \begin{bmatrix}
I\\ 0\\ 0
\end{bmatrix} & \begin{bmatrix}
0\\ I\\ 0
\end{bmatrix} & \begin{bmatrix}
0\\ 0\\ 0
\end{bmatrix} &\begin{bmatrix}
0\\ 0\\ I
\end{bmatrix}\\ \hline
0 & 0 & 0 & I\\ C_i & 0 & 0 & D_i
\end{bmatrix}<0
\end{align}
    \end{figure*}

\section{Distributed controller synthesis from data}
So far we have considered the performance analysis of (interconnected) systems from data for the channel $u \to y$. We will now consider a distributed control problem for the interconnected system \eqref{eq:i}, where we take $u_i$ and $y_i$ as the control input and measured output respectively. Recall that we assume that input-state data is collected for to determine $\Sigma_\mathcal{D}^i$ for each $i$. With the system matrices defined as $C_i=I$, $D_i=0$, this implies only state-measurements are available for control. We note, however, that $C_i$ is allowed to be chosen arbitrarily in this section and that $D_i=0$; this implies that output measurements only can be utilized for the controller implementation. Future research will focus on extending the framework to the case when only input-output data is available for synthesis. The problem under consideration is to guarantee that the channel $w\to z$ achieves $\mathcal{H}_\infty$ performance $\gamma>0$, with performance output
\begin{align} \label{eq:perf}
z_i=C_i^zx_i+\sum_{j\in\mathcal{N}_i}C_{ij}^zx_j+D_i^zu_i.
\end{align}

We consider a distributed controller that is an interconnected system with dynamic subsystems
\begin{align} \label{eq:conti}
\begin{bmatrix}
\xi_i(k+1)\\ o_i(k)\\ u_i(k)
\end{bmatrix}=\Theta_i\begin{bmatrix}
\xi_i(k)\\ s_i(k)\\ y_i(k)
\end{bmatrix},\quad i=1,\dots,L,
\end{align}
where $\xi_i\in\mathbb{R}^{n_i}$ is the state of controller $i$ and $o_i=\operatorname{col}_{j\in\mathcal{N}_i} o_{ij}$, $s_i=\operatorname{col}_{j\in\mathcal{N}_i} s_{ij}$ are interconnection variables satisfying $s_{ij}=o_{ji}\in\mathbb{R}^{n_{ij}}$ for $(i,j)\in\mathcal{E}$.

\begin{proposition}[Distributed control from data]\label{prop:existci}
Let $\Psi_i$ and $\Phi_i$ be matrices that are a basis of $\operatorname{ker}\begin{bmatrix} C_i & 0\end{bmatrix}$ and $\operatorname{ker}\begin{bmatrix}
0 & I &(D_i^z)^\top
\end{bmatrix}$, respectively, and let $n_{ij}=3n_i$. If there exist $P_i$, $\bar{P}_i$, $Z_i$, $\bar{Z}_i$, $\alpha_i$ such that $P_i>0$, $\bar{P}_i>0$, $\alpha_i>0$ \eqref{eq:datLMIdistr1}, \eqref{eq:datLMIdistr2} hold (see next page) with $\beta_i=\alpha_i^{-1}$ and
\begin{align*}
\begin{bmatrix}
P_i & I\\ I & \bar{P}_i
\end{bmatrix}\geq 0,
\end{align*}
then there exist $\Theta_i$, $i\in\mathcal{V}$, so that all closed-loop interconnected systems described by \eqref{eq:i}, \eqref{eq:perf} and \eqref{eq:conti} with subsystems $(A_i,A_{\mathcal{N}_i},B_i)\in \Sigma_\mathcal{D}^i$ achieve $\mathcal{H}_\infty$ performance $\gamma$.
\end{proposition}

\begin{figure*}[b]
    \hrulefill
\begin{align} 
\Psi_i^\top(\star)^\top\left[\begin{array}{cc|cc|cc|cc}
-P_i & 0 & 0 & 0 & 0 & 0 & 0 & 0\\
0 & P_i & 0 & 0& 0 & 0 & 0 & 0\\ \hline
0 & 0 & Z_i^{11}& Z_i^{12} & 0 & 0 & 0 & 0\\
0 & 0 & (Z_i^{12})^\top & Z_i^{22} & 0& 0 & 0 & 0\\ \hline
0 & 0 & 0 & 0 & -\alpha_i R_\mathcal{D}^i & -\alpha_i(S_\mathcal{D}^i)^\top & 0 & 0\\
0 & 0 & 0 & 0 & -\alpha_i S_\mathcal{D}^i & -\alpha_i Q_\mathcal{D}^i & 0 & 0\\ \hline
0 & 0 & 0 & 0  & 0 & 0 & -\gamma^2 I & 0\\
0 & 0 & 0 & 0  & 0 & 0 & 0 & I
\end{array}\right] \begin{bmatrix}
I & 0 & 0 & 0\\
0 & 0 & I & I\\ \hline
\mathbf{1}\otimes I & 0 & 0 & 0\\ 0 & I & 0 & 0\\ \hline
0 & 0 & I & 0\\ \begin{bmatrix}
I\\ 0\\ 0
\end{bmatrix} & \begin{bmatrix}
0\\ I\\ 0
\end{bmatrix} & \begin{bmatrix}
0\\ 0\\ 0
\end{bmatrix} &\begin{bmatrix}
0\\ 0\\ 0
\end{bmatrix}\\ \hline
0 & 0 & 0 & I\\ C_i^z & C_{\mathcal{N}_i}^z & 0 & 0
\end{bmatrix}\Psi_i&<0 \label{eq:datLMIdistr1}\\
\Phi_i^\top (\star)^\top\left[\begin{array}{cc|cc|cc|cc}
-\bar{P}_i & 0 & 0 & 0 & 0 & 0 & 0 & 0\\
0 & \bar{P}_i & 0 & 0& 0 & 0 & 0 & 0\\ \hline
0 & 0 & \bar{Z}_i^{11}& \bar{Z}_i^{12} & 0 & 0 & 0 & 0\\
0 & 0 & (\bar{Z}_i^{12})^\top & \bar{Z}_i^{22} & 0& 0 & 0 & 0\\ \hline
0 & 0 & 0 & 0 & -\beta_i \bar{R}_\mathcal{D}^i & -\beta_i (\bar{S}_\mathcal{D}^i)^\top & 0 & 0\\
0 & 0 & 0 & 0 & -\beta_i \bar{S}_\mathcal{D}^i & -\beta_i \bar{Q}_\mathcal{D}^i & 0 & 0\\ \hline
0 & 0 & 0 & 0  & 0 & 0 & -\gamma^{-2} I & 0\\
0 & 0 & 0 & 0  & 0 & 0 & 0 &  I
\end{array}\right] \begin{bmatrix}
I & 0 & 0 & 0\\
0 & 0 & I & I\\ \hline
\mathbf{1}\otimes I & 0 & 0 & 0\\ 0 & I & 0 & 0\\ \hline
0 & 0 & I & 0\\ \begin{bmatrix}
I\\ 0\\ 0
\end{bmatrix} & \begin{bmatrix}
0\\ I\\ 0
\end{bmatrix} & \begin{bmatrix}
0\\ 0\\ 0
\end{bmatrix} &\begin{bmatrix}
0\\ 0\\ 0
\end{bmatrix}\\ \hline
0 & 0 & 0 & I\\ C_i^z & C_{\mathcal{N}_i}^z & 0 & 0
\end{bmatrix}_\perp\!\!\!\! \Phi_i&>0\label{eq:datLMIdistr2}
\end{align}
    \end{figure*}

\begin{remark}
The conditions in Proposition~\ref{prop:existci} are sufficient for any choice of $\alpha_i$, e.g. $\alpha_i:=\alpha=1$ for all $i$, and the conditions are LMIs for fixed $\alpha_i$.  Conservatism can be reduced by, e.g., verifying feasibility of the LMIs on a discrete interval for $\alpha$.
\end{remark}

In particular, Proposition~\ref{prop:existci} implies that the existence of a distributed controller for which the `true' interconnected system achieves $\mathcal{H}_\infty$ performance, can be verified by checking a set of LMIs based on noisy input-state data. Suitable matrices $P_i$, $\bar{P}_i$, $Z_i$, $\bar{Z}_i$ are thus indirectly based on the data; these matrices can be used for the subsequent construction of the controller matrices $\Theta_i$ as described in \cite{steentjes2021}, cf. \cite{langbortetal2004}, \cite{horssen2016}. We remark that neither the existence, nor the construction of $\Theta_i$ is based on the unknown matrices $(A_i,A_{\mathcal{N}_i},B_i)$.

\section{Examples}
\subsection{Example~1: $\mathcal{H}_\infty$-norm analysis from noisy data}
Consider a system of the form \eqref{eq:sys0} with $L=3$,
\begin{align*}
A_0=\begin{bmatrix}
0.5 & 0.1 & 0\\ 0.1 & 0.4 & 0.1\\ 0 & 0.1 & 0.6
\end{bmatrix}\quad \text{ and }\quad B_0=I.
\end{align*}
We choose $y=x$ so that $C=I$ and $D=0$. The input entries are drawn from a normal distribution with zero mean and unit variance. The noise $w(k)$ is drawn uniformly from the set $\{w\,|\, \|w\|_2\leq \sigma\}$, where $\sigma>0$ determines the noise level. Hence, considering the set $\mathcal{W}$ with $Q_w=-I$, $S_w=0$ and $R_w =N\sigma^2 I$, we have that the noise sequence satisfies $W_-\in\mathcal{W}$.

\begin{figure}[!t]
\centering
\includegraphics[width=3.5in]{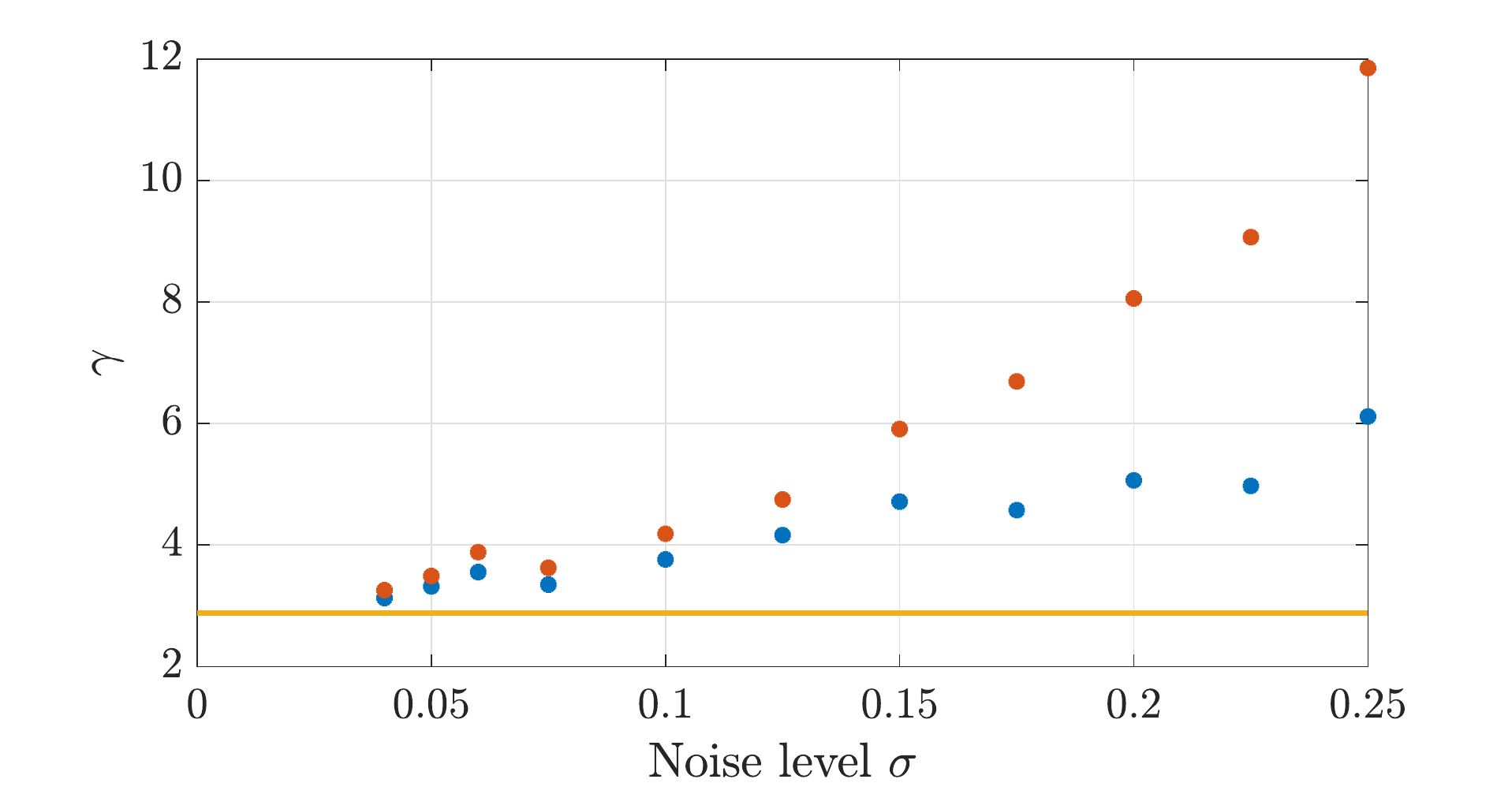}
\caption{Example~1: Upper bound on the $\mathcal{H}_\infty$ norm determined from noisy data with increasing noise levels $\sigma$ via lumped (blue) and structured (red) data analysis. The $\mathcal{H}_\infty$ norm of the true system is shown in orange.}
\label{fig:noisevsnorm}
\end{figure}

The aim is to find an upperbound on the $\mathcal{H}_\infty$ norm of the channel $u\to y$ using the noisy data $(U_-,X)$ with $N=50$ samples. The true $\mathcal{H}_\infty$ norm is $\gamma_0= 2.8836$. We choose eleven noise levels $\sigma$ in the interval $[0.04,0.25]$ and generate one data set for each noise level. For each data set, we minimize $\gamma$ subject to \eqref{eq:datLMI} with $Q = \gamma^2 $, $S=0$ and $R=-I$. The results are displayed in Figure~\ref{fig:noisevsnorm} in blue. By Proposition~\ref{prop:disdata}, the corresponding solutions satisfy \eqref{eq:LMI}, hence $\gamma$ is an upperbound on the $\mathcal{H}_\infty$ norm for all systems in $\Sigma_\mathcal{D}$ \cite{schererweilandLMI} and, therefore, for $(A_0,B_0)$.

Next, we perform the analysis through Proposition~\ref{prop:disdatai} using the same data sets. It is clear that $W_i^-\in\mathcal{W}_i$ for each $i$ with $Q_w^i=-I$, $S_w^i=0$ and $R_w^i =N\sigma^2 I$. For each data set, we minimize $\gamma$ subject to the LMIs \eqref{eq:datLMIint} for $i=1,2,3$. The resulting values of $\gamma$ provide a guaranteed upper bound on the $\mathcal{H}_\infty$ norm of $u\to y$ and are shown in Figure~\ref{fig:noisevsnorm} in red.

The computed value of $\gamma$ using either Proposition~\ref{prop:disdata} or Proposition~\ref{prop:disdatai} is a guaranteed upper bound for the $\mathcal{H}_\infty$ norm of the true system for all noise levels. The bound provides a good approximation of $\gamma_0$ for low noise levels. For increasing noise levels, the bound $\gamma$ becomes more conservative for both methods. Comparing the results from Proposition~\ref{prop:disdata} (unstructured data) with Proposition~\ref{prop:disdatai} (structured data), the bounds obtained from \eqref{eq:datLMIint} are conservative with respect to those from \eqref{eq:datLMI} for higher noise levels, while the difference is small for low noise levels. By solving the unstructured data-based conditions in \cite[Theorem~4]{koch2020provably}, we find the same bounds as obtained per Proposition~\ref{prop:disdata}, as expected from the duality of the results.

\subsection{Example~2: Distributed $\mathcal{H}_\infty$ controller synthesis from noisy data}
Consider an interconnected system with $L=25$ subsystems, each having one state ($n_i=1$). The subsystems are interconnected according to a \emph{cycle} graph $\mathcal{G}$ and the matrices $A_i$ and $A_{ij}$ are drawn uniformly on the interval $[0,1]$ and $[0,0.1]$, respectively, and $B_i=1$. We consider $y_i=x_i$ for all subsystems and consider the performance output $z_i=x_i$, so that $C_i^z=I$ and $C_{ij}^z=D_i^z=0$. For the data acquisition, the input entries are drawn from a normal distribution with zero mean and unit variance. The noise signals $w_i(k)$ are drawn uniformly from the set $\{w\,|\, |w|\leq \sigma\}$, where $\sigma = 0.05$ is the noise level. Hence, considering the sets $\mathcal{W}_i$ with $Q_w^i=-I$, $S_w^i=0$ and $R_w^i =N\sigma^2 I$, we have that the noise sequences satisfy $W_-^i\in\mathcal{W}_i$, $i=1,\dots,L$.

The goal is to synthesize a distributed controller that yields an upperbound $\gamma$ on the $\mathcal{H}_\infty$ norm of the channel $w\to z$, without using knowledge of $A_i$, $A_{ij}$ and $B_i$. First, we verify what the smallest upperbound $\gamma$ is, for which there exists a  \emph{model-based} distributed controller by the \emph{nominal} LMIs in \cite[Theorem~2]{horssen2016}. This smallest upperbound of $\gamma$ is $1.00$ and serves as a benchmark: our data-based method for distributed control cannot perform better than the model-based distributed controller. We generate the data matrices $(U_i^-,X_i)$ for $N=50$ samples. For $\alpha_i:=\alpha=1$, we observe that the LMIs \eqref{eq:datLMIdistr1} and \eqref{eq:datLMIdistr2} are feasible for $\gamma =1.10$. Hence, by Proposition~\ref{prop:existci}, there exists a distributed controller that achieves an $\mathcal{H}_\infty$ norm less than $1.10$ in closed-loop with the true interconnected system.

\begin{figure}[!t]
\centering
\includegraphics[width=3.5in]{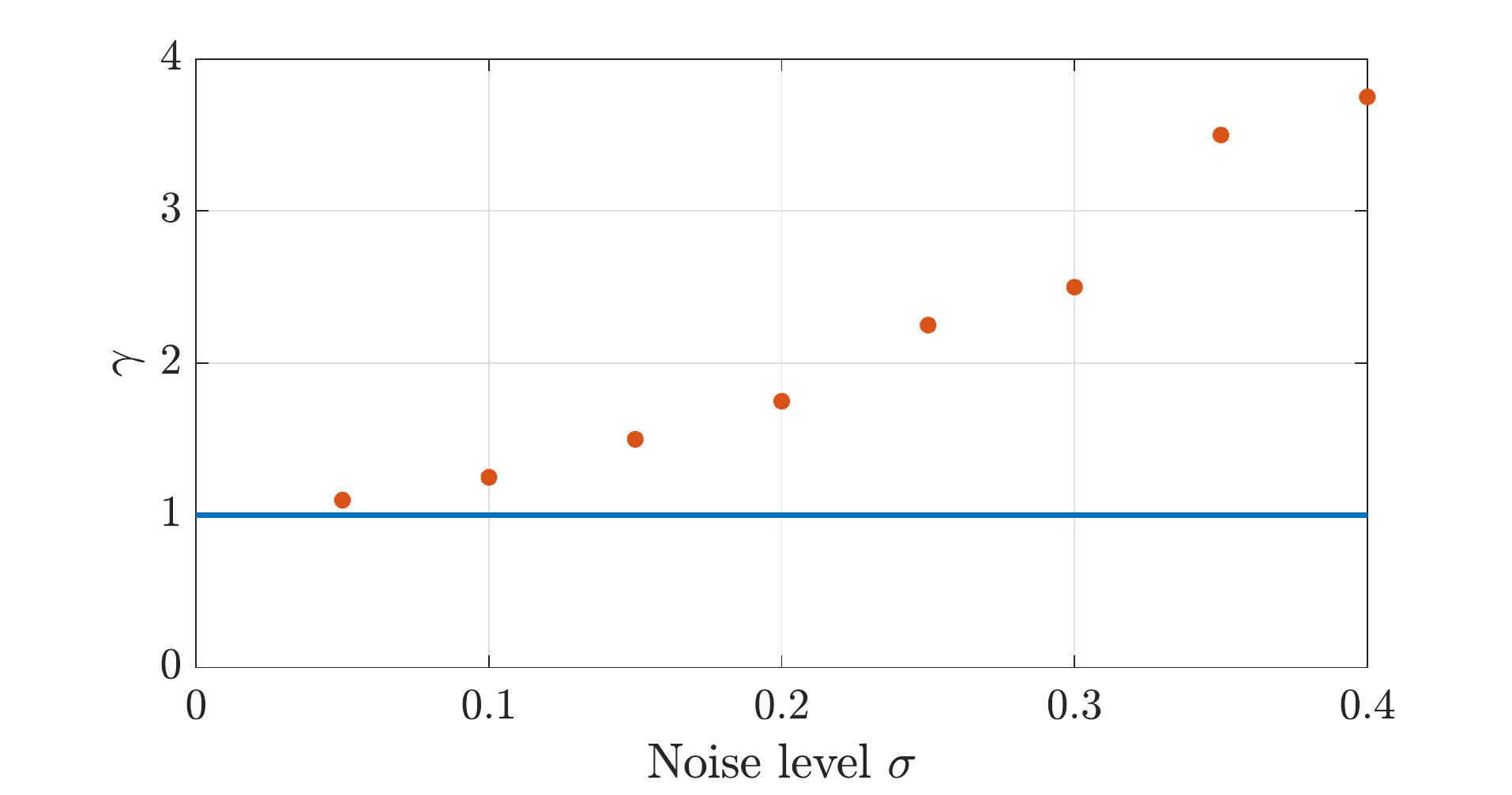}
\caption{Example~2: Achievable $\mathcal{H}_\infty$ norm with distributed control from noisy data with increasing noise levels $\sigma$ (red) and achievable $\mathcal{H}_\infty$ norm with distributed control computed using the true system (blue).}
\label{fig:noisevsnormsyn}
\end{figure}

Next, we increase the noise level, up to $\sigma = 0.4$. The resulting values of $\gamma$ are shown in Figure~\ref{fig:noisevsnormsyn} and obtained from the conditions in Proposition~\ref{prop:existci} by varying $\alpha$ in a discrete interval. We observe that the conservatism increases for increasing noise levels. This can be explained by the fact that the size of the sets $\Sigma_\mathcal{D}^i$ increases, leading to the existence of a more conservative distributed controller that achieves $\mathcal{H}_\infty$ performance for all interconnected systems compatible with the data.

\section{Concluding remarks}
We have considered the problem of analyzing the $\mathcal{H}_\infty$ norm of an interconnected system and finding a distributed controller that achieves $\mathcal{H}_\infty$ performance based on noisy data. First, we considered a dual parameterization of the set of systems that are compatible with the data with respect to the parameterization developed in~\cite{koch2020provably} and we presented a dual result for data-based dissipativity analysis. The introduction of dual parameterizations of data-compatible subsystems, allowed us to introduce an interconnected system with LFT representations of the subsystems. We have presented sufficient LMI conditions based on data that guarantee $\mathcal{H}_\infty$ performance or the existence of a distributed controller that achieves $\mathcal{H}_\infty$ performance. The LMI conditions are sufficient as in~\cite{koch2020provably}, unlike the necessary and sufficient conditions in~\cite{vanwaarde2020noisy}. Nonetheless, to the best of the authors' knowledge, the developed framework provides the first scalable method to data-based performance analysis and distributed controller synthesis with guaranteed $\mathcal{H}_\infty$ performance from noisy data.

\appendices
\section{Proof of Lemma~\ref{lem:dualparam}} \label{app:lemproof}
\begin{proof}
Let $M:=\begin{bmatrix}
A & B
\end{bmatrix}$ and
\begin{align*}
\bar{P}:=\begin{bmatrix}
\bar{Q}_{\mathcal{D}} & \bar{S}_{\mathcal{D}}\\ \bar{S}_{\mathcal{D}}^\top & \bar{R}_{\mathcal{D}}
\end{bmatrix}\in\mathbb{R}^{n\times n}.
\end{align*}
Since $\operatorname{col}(U_-,X_-)$ has full row rank, $(A,B)\in\Sigma_\mathcal{D}$ if and only if
\begin{align*}
\begin{bmatrix}
-M^\top \\ I
\end{bmatrix}^\top \begin{bmatrix}
\bar{Q}_{\mathcal{D}} & \bar{S}_{\mathcal{D}}\\ \bar{S}_{\mathcal{D}}^\top & \bar{R}_{\mathcal{D}}
\end{bmatrix}\begin{bmatrix}
-M^\top\\ I
\end{bmatrix}\geq 0 \quad \text{ and } \quad \bar{Q}_\mathcal{D}<0,
\end{align*}
by Lemma~\ref{lem:param}. This is equivalent with
\begin{align} \label{eq:eqparam}
\bar{P}\geq 0 \text{ on } \operatorname{im}\begin{bmatrix}
-M^\top\\ I
\end{bmatrix}\quad \text{ and } \quad \bar{P}<0 \text{ on } \operatorname{im}\begin{bmatrix}
I\\ 0
\end{bmatrix}.
\end{align}
Since the direct sum of $\operatorname{im}\begin{bmatrix}
-M^\top\\ I
\end{bmatrix}$ and $\operatorname{im}\begin{bmatrix}
I\\ 0
\end{bmatrix}$ is equal to $\mathbb{R}^n$, it follows by the dualization lemma~ \cite[Lemma~4.9]{schererweilandLMI} that~\eqref{eq:eqparam} holds if an only if
\begin{align*}
\bar{P}^{-1}\leq 0 \text{ on } \operatorname{im}\begin{bmatrix}
-M^\top\\ I
\end{bmatrix}^\perp\quad \text{ and } \quad \bar{P}^{-1}>0 \text{ on } \operatorname{im}\begin{bmatrix}
I\\ 0
\end{bmatrix}^\perp\!\!,
\end{align*}
which is equivalent with
\begin{align*}
\bar{P}^{-1}\leq 0 \text{ on } \operatorname{im}\begin{bmatrix}
I\\ M
\end{bmatrix}\quad \text{ and } \quad \bar{P}^{-1}>0 \text{ on } \operatorname{im}\begin{bmatrix}
0\\ I
\end{bmatrix}.
\end{align*}
Thus, \eqref{eq:eqparam} holds if and only if
\begin{align*}
\begin{bmatrix}
I\\ M
\end{bmatrix}^\top \begin{bmatrix}
Q_\mathcal{D} & S_\mathcal{D}\\ S_\mathcal{D}^\top & R_\mathcal{D}
\end{bmatrix}\begin{bmatrix}
I\\ M
\end{bmatrix}\leq 0\quad \text{ and } \quad R>0,
\end{align*}
which proves the assertion.
\end{proof}

\section{Definition scales}\label{app:scales}
Let $X_{ij}^{11}$ and $\bar{X}_{ij}^{11}$ be symmetric matrices. We define
\begin{align*}
Z_i^{11}&:=-\operatorname*{diag}_{j\in\mathbb{Z}_{[1:L]}} X_{ij}^{11}, Z_i^{22}:=\operatorname*{diag}_{j\in\mathbb{Z}_{[1:L]}} X_{ji}^{11},\\
\bar{Z}_i^{11}&:=-\operatorname*{diag}_{j\in\mathbb{Z}_{[1:L]}} \bar{X}_{ij}^{11}, \bar{Z}_i^{22}:=\operatorname*{diag}_{j\in\mathbb{Z}_{[1:L]}} \bar{X}_{ji}^{11},\\
Z_i^{12}&:= \operatorname{diag}\left( -\operatorname*{diag}_{j\in\mathbb{Z}_{[1:i]}} X_{ij}^{12},\operatorname*{diag}_{j\in\mathbb{Z}_{[i+1:L]}} (X_{ji}^{12})^\top \right),\\
\bar{Z}_i^{12}&:= \operatorname{diag}\left( -\operatorname*{diag}_{j\in\mathbb{Z}_{[1:i]}} \bar{X}_{ij}^{12},\operatorname*{diag}_{j\in\mathbb{Z}_{[i+1:L]}} (\bar{X}_{ji}^{12})^\top \right).
\end{align*}

\bibliographystyle{IEEEtran}
\bibliography{../rfrncs20}

\end{document}